\documentclass[runningheads]{svmult}

\usepackage{makeidx}   
\usepackage{graphicx}  
\usepackage{subeqnar}  
\usepackage{multicol}  
\usepackage{physprbb}  
\makeindex             

\newcommand{\mic}{$\mu$m}

\begin{document}
\title*{Multiple Integral Field Spectroscopy}
\toctitle{Multiple Integral Field Spectroscopy}
\titlerunning{Multiple Integral Field Spectroscopy}

%
%
%
%
\author{Gillian Wright\inst{1}
\and Rob Ivison\inst{1} \and Peter Hastings\inst{1} \and Martyn Wells\inst{1}
\and Ray Sharples\inst{2} \and Jeremy Allington-Smith\inst{2} \and Robert Content \inst{2}
}
\authorrunning{Gillian Wright et al.}

\institute{
UK Astronomy Technology Centre, Royal Observatory, Edinburgh EH9 3HJ, UK
\and
University Of Durham, Dept.\ of Physics, South Road, Durham DH1 3LE, UK
}

\maketitle

\begin{abstract}
Integral-field spectroscopy is the most effective method of exploiting
the superb image quality of the ESO--VLT, allowing complex
astrophysical processes to be probed on the angular scales currently
accessible only for imaging data, but with the addition of information
in the spectral dimension. We discuss science drivers and requirements
for multiple deployable integral fields for spectroscopy in the
near-infrared.  We then describe a fully modular instrument concept
which can achieve such a capability over a 5--10$'$ field with up to
32 deployable integral fields, each fully cryogenic with 1--2.5\,\mic\
coverage at a spectral resolution of $\sim$3000, each with a 4$''$
$\times$ 4$''$ field of view sampled at 0.2\,arcsec\,pixel$^{-1}$ to
take advantage of the best $K$-band seeing.
\end{abstract}

\section{Introduction}

The superb image quality of the ESO--VLT makes possible the study of
spatial structure in unprecedented detail. Integral-field-unit (IFU)
spectroscopy is a very powerful technique, allowing complex processes,
both physical, chemical and kinematic, be probed with the same angular
resolution as with conventional imaging data. As well as the
ability to take spectra of contiguous areas, the now well-known
advantages of IFU spectroscopy also include the elimination of slit
losses and the relative ease of target acquisition. In the
near-infrared, it is key to understanding the distribution, excitation
and kinematics of gas and/or stars in objects ranging from obscured
protostellar complexes to morphologically peculiar high-redshift galaxies.

Since many important advances will come from the study of statistical
samples of faint astrophysical objects, simultaneous integral-field
spectroscopy of several sources in a single field will greatly
increase the efficient use of the precious VLT resource.  The 2k
$\times$ 2k pixels available in foreseeable infrared arrays means that
fully covering the focal plane of an 8m-class telescope with an IFU
capable of fine spatial sampling will remain impracticable for many
years. To allow high-spatial-resolution integral-field spectroscopy of
several targets within a field, we therefore need the ability to
position a number of small-field IFUs anywhere within a larger patrol
field, a non-trivial task.

The capability to work at wavelengths longer than 1.7\,\mic, where
thermal emission dominates the background, is crucial. For example,
the study of H\,$\alpha$ emission in galaxies at $z > 1.8$, or using
CO, He\,{\sc i} and He\,{\sc ii} lines to probe heavily obscured
proto-stars all require integral-field spectroscopy with $\lambda >
2$\,\mic. The challenges of cryogenically cooling the optics and the
deployment mechanism therefore have to be solved.  We have developed a
concept for a modular instrument with deployable optics to position 32
IFU fields over a 10$'$ field.  We summarise the science drivers to
estimate the required number of IFU fields and present strawman
functional requirements for the instrument. Finally, we then describe
the results of our design study.

\section{Science Drivers}

Extremely deep optical and infrared surveys --- the Hubble Deep Fields
(Williams et al.\ 1996) and their ilk --- are revolutionising the
study of galaxy evolution by providing large samples of galaxies at a
range of redshifts for which detailed comparison of the distribution,
excitation and kinematics of the gas and stars with nearby analogues
may be made. Radio, X-ray and submm surveys (using SCUBA, MAMBO, VLA,
{\em XMM} and {\em Chandra}, and future missions such as BLAST, {\em
SIRTF} and {\em Herschel}) are providing large, complementary samples
of X-ray- and radio-loud AGN and distant starbursts, many severely
obscured by dust (e.g.\ Smail et al.\ 1997; Richards 2000). $K$-band surveys find around $\sim10^4$ $\rightarrow$
$\sim10^5$\,degree$^{-2}$ galaxies to $K\sim$ 19.5 $\rightarrow$ 22
(Huang et al.\ 2001). A fraction of these typically fulfil the
selection criteria for an astrophysical application so that a
programme will typically select a few tens of objects distributed over
a 5--10$'$ field (see Table~1). 

\begin{figure}[b]
\begin{center}
\includegraphics[width=.6\textwidth]{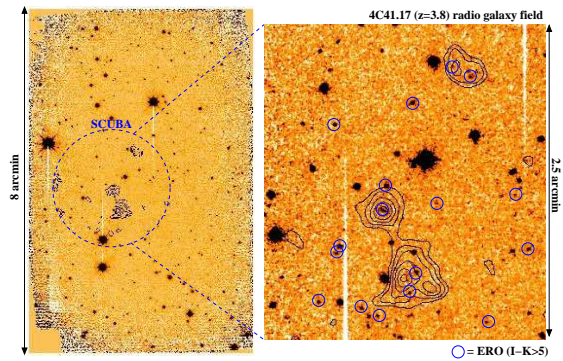}
\end{center}
\caption[]{The field around the $z=3.8$ radio galaxy, 4C\,41.17, as imaged
in $K$, with SCUBA 850-$\mu$m contours superimposed over a small region
(Ivison et al.\ 2000). EROs with $I-K>5$, some associated with the
SCUBA galaxies, are circled.}
\end{figure}

One of the defining characteristics of highly-obscured AGN and
star-forming galaxies is their red colours in the rest-frame optical
and near-infrared. By their very nature these galaxies are difficult
to identify in optical surveys. A few
examples of such extremely red objects (EROs --- Hu \& Ridgway 1994)
turned up through deep near-infrared surveys, but these were viewed as
little more than curiosities. A number of recent developments have
forced a reassessment of this situation and it now appears that
highly-obscured sources may host
power sources which dominate the extragalactic background light in the 
hard X-ray and submm bands (Smail et al.\ 1999).
Under half of the ERO population has been identified with obscured
starbursts, the rest with massive elliptical galaxies at $z\sim$ 1--2 whose red
colours arise from their evolved stellar populations. The long-term
goals of these studies are to constrain the early formation of massive
galaxies at high redshifts and to derive the relative histories of
obscured star formation and accretion activity to test models of
galaxy and black hole evolution. 
Knowledge of the redshift distribution is crucial -- $N(z)$ breaks
degeneracies in the models and allows the mass and luminosity of the
sources to be explored, e.g.\ via observations of CO. Near-infrared
IFU spectroscopy can also provide extinction estimates, independent
star-formation rates, and probe for signs of AGN activity (e.g.\ broad
H\,$\alpha$ emission, Ivison et al.\ 2000).

At $K\sim 21$ the surface density of $I-K>5$ EROs is
$\sim$360\,degree$^{-2}$, and so we immediately see the benefit of
spectroscopic multiplexing. The surface density of submm galaxies
above 2\,mJy at 850$\mu$m is $\sim 4000$\,degree$^{-2}$ with at least
half expected to be associated with EROs. A multiplexing factor of
$\sim$30 may thus be appropriate to take advantage of upcoming
surveys. It is difficult, however, to predict this factor precisely,
which is an excellent reason to build a modular instrument (\S3.3).

Studies of galaxy evolution ultimately require full two-dimensional
spectroscopy in the $J$, $H$ and $K$ bands for meaningful samples of
sources, to cover the key rest-frame optical diagnostic lines at
$z>0.5$. Virtually all galaxies are spatially resolved in the best VLT
$K$-band seeing, and many will have complex morphologies --- the
result of interactions and mergers for the obscured systems, the
result of rotation for some of the field galaxy population.
Integral-field spectroscopy will be essential to determine their
detailed internal properties.

Other obvious astrophysical applications for near-infrared
multiple-IFU spectroscopy include investigating the build-up of spiral
disks, exploiting gravitational amplification of distant galaxies by
foreground cluster lenses probing and investigating young globulars in
starburst galaxies.

\begin{table}
\caption{Estimated source densities.}
\begin{center}
\begin{tabular}{lll}
\hline\noalign{\smallskip}
Object class & Surface Density /arcmin$^{-2}$  \\
\noalign{\smallskip}
\hline
\noalign{\smallskip}
All galaxies (K $<$ 20) & 10         \\
HDF irregulars ($z$ = 0.5--1)& 5-- \\
Emission-line galaxies ($z$ = 2--2.7) & 2  \\
Gravitational arcs in clusters & 1 \\
EROs to $K\sim 21$ & 0.1 \\
Super-star clusters & 1-10 \\
Prolyds/microjets in Orion & 1  \\
T-Tauri stars & 0.1  \\
\hline
\end{tabular}
\end{center}
\label{Tab1a}
\end{table}

Our instrument concept can provide a maximum number of pick-offs of
about 32.  To use all 32 pick-offs, the mean surface density of
randomly-distributed sources needs to be 0.5--2\,arcmin$^{-2}$ for a
patrol field of 5--10\,arcmin, i.e.\ about 40 per field. This is
because the pick-off mechanism inevitably vignettes a small fraction
of the field. From Table~1 it is obvious that about 30 deployable IFUs
are needed for the science --- there is no strong case for hundreds of
simultaneous measurements, and too few IFUs would not provide optimum
efficiency. 

\subsection{Proposed instrument capabilities}

Based on the science drivers, we concluded that the instrument should
have the following capabilities: \begin{itemize}
\item
1--2.5\,\mic\ coverage, fully cryogenic for good $K$-band performance;

\item
patrol field of 5--10\,arcmin diameter;

\item
approximately 30 individual deployable fields;

\item
spatial sampling in the IFUs should be 0.2\,arcsec\,pixel$^{-1}$ to
capitalise on the best seeing conditions at $K$;

\item
field of an IFU should be around 4 $\times$ 4\,arcsec so that for most
sources some sky area is included;

\item
a spectral resolution of $R\sim$ 3000 to work effectively between OH lines;

\item
2k $\times$ 2k detectors were assumed, with up to 8 needed for a full
set of 32 deployable units;

\item
the deployment mechanism must be capable of positioning an IFU to
within 1/5 of an IFU field.  Since the field of the IFU and the size of
the array mean that there is no cross dispersion, if accurate
relative astrometry is needed across the full spectral range then the
$JHK$ spectra must be obtained before reconfiguring the field.
\end{itemize}

\section{Instrument concept}

There are two basic approaches to deploying reflective image slicers
--- physically move them in the focal plane, or tilt/move a pick-off
mirror to direct the desired field into a fixed image slicer. Moving
the IFU means that the position of the spectra on the detector will
vary depending on the position of the IFU in the field, and the design
of a cryogenic mechanism which does not obstruct the moving modules is
very difficult. A simpler method is to use a pick-off mirror system to
steer the desired field area into a fixed array of image slicers. This
also has the advantage of producing a fixed spectrum on the
detector(s).  Additional goals for the design of the pick-off
mechanism were to use proven technologies, stepper motors and gears
for the moving parts, and so a small pick-off mirror is preferred to a
heavier image slicer.

\subsection{Pick-off arms}

\begin{figure}[b]
\begin{center}
\includegraphics[width=.7\textwidth]{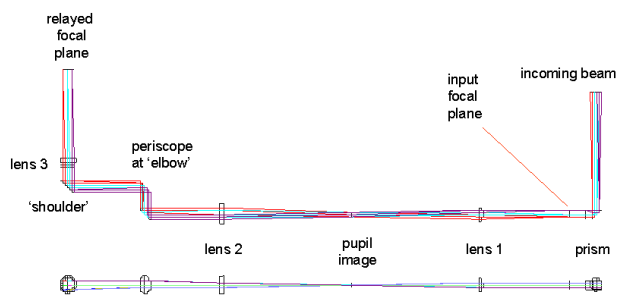}
\end{center}
\caption[]{Optical design of the pick-off arm.
The light is intercepted just ahead of the focal plane and
folded inside a tubular arm by a small prism. The beam is
then collimated by an achromatic lens, producing a real pupil at a cold
stop. The light is folded twice to form a periscope, allowing the
arm, lens, stop and prism to rotate about an axis parallel to the
incoming beam. The light passes along a shorter arm and is again
folded to lie parallel to the original input beam. Rotations about the
output axis and the periscope axis are used to position the prism to
capture the desired field of view for the IFU.}
\label{eps1}
\end{figure}

An articulated arm carries relay optics to feed small portions of the
focal plane to each IFU which are at fixed locations in the instrument. 
The focal plane (where the
pick-offs move) should ideally be flat to remove the need for a
focus mechanism for each and every pick-off. It is also telecentric,
which avoids pupil motion dependent on field position.
Field correction is achieved by lenses that form the cryostat window. The
principle of operation of the pick-off arm concept is shown in
Figure~2.  A potential difficulty of using steerable mirrors --- the need to
compensate for changes in the optical path length between the focal
plane and the spectrograph --- is avoided because a collimated beam is
passed around the elbow and shoulder and so no other mechanisms are
required.  

\begin{figure}[b]
\begin{center}
\includegraphics[width=.6\textwidth]{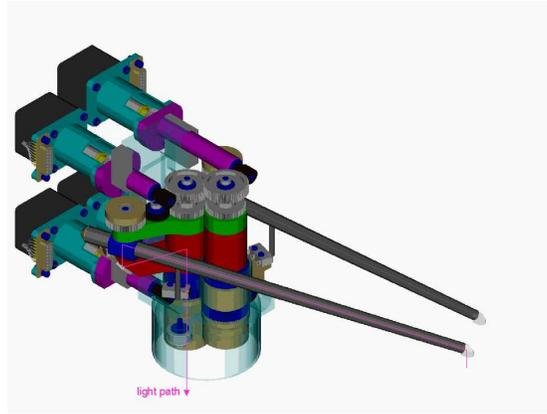}
\end{center}
\caption[]{Mechanal design of the arm.
The main arm is a tapered tube of carbon-fibre-reinforced plastic
which carries the collecting prism at its tip and contains the lens
and stop, counterbalanced about the elbow axis and driven by
a sector attached above the arm which carries the bearings. 
At the elbow, the light is
reflected downwards into another arm which contains the lower
mirror of the elbow periscope and the final shoulder mirror. This arm
carries no loads other than the weight of the mirrors. The light
passes to the output focal plane along the tubular main shaft of the
shoulder axis.}
\label{eps1}
\end{figure}

Although simple in theory, the articulated arm has a number
of complexities in its engineering: it must be light and stiff,
compact and protected against control failure.  The end result, shown
in Figure~3, is therefore quite complex.

Both the necessary rotations are produced by stepping motors
which drive worm and wheel mechanisms.  In the event of a 
malfunction in the control chain,
a stopwork prevents the shoulder from turning more than
$\pm$55$^{\circ}$ by stalling the motor. The elbow rotation has two
stages of gear reduction --- an initial worm and wheel identical to (and
concentric with) the shoulder drive, followed by a gear and sector
reduction to give a fine motion to the prism at the far end of the
main arm. An additional benefit is that the weight of the elbow drive
motor is not carried by the shoulder bearings or any of the moving
support structures. This layout does mean, however, that a rotation of
the shoulder induces a rotation of the elbow. To keep the main arm
parallel to its original orientation as the shoulder moves, both
shoulder and elbow drives must operate.
If the elbow drive fails for any reason, it must be possible to drive
the shoulder (if only to retract the arm from the field of view)
without damaging the elbow joint. Consequently, the main arm is
mounted so that it is driven positively in one direction but via a
spring in the other direction. This gives the freedom to deal with
such a failure of the elbow motor, or a clash between an arm and its
neighbours. A stopwork similar to that on the shoulder drive prevents
excessive rotation of the elbow.

\subsection{Packaging the relay arms with IFUs and spectrographs}

\begin{figure}[b]
\begin{center}
\includegraphics[width=.7\textwidth]{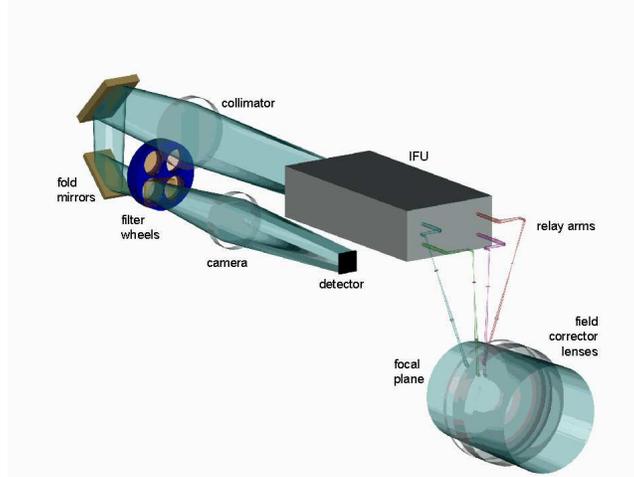}
\end{center}
\caption[]{Unit spectrograph concept.}
\label{eps1}
\end{figure}

Rather than
building a single large spectrograph, the instrument is instead
designed to be highly modular with a number of identical spectrographs
each fed by only a small number of IFUs. There are several variations
possible, ranging from 1 IFU --- 1 spectrograph --- 1 array to 4 IFUs
--- 1 spectrograph --- 1 array.  The selection of one these options
will ultimately depend on the number of arrays available and a more
detailed science trade-off between the IFU fields and number of
spectral elements. Since much of the mass and space envelope for the
instrument is occupied by the IFU deployment mechanisms, the
spectrograph modules must be compact and simple.
The maxiumum number of pick-off fields of view is provided by the
solution in which 4 integral fields feed 1 spectrograph, as shown in Figure~4.
Small image slicers such as those described in Content et
al.\ 1997 or Wells et al.\
2000 are packaged in groups of four
to provide a compact unit at the end of the feed arms 
An optical design has been developed for a compact 2 mirror
plus 1 grating spectrograph that has 2 slits at the entrance aperture.
Each slit is covers 2k pixels and is formed by two of the four image
slicers; the spectra cover 1k pixels.  If only half the arms are used,
it would be possible in principal to obtain spectra covering 2k
pixels.

\subsection{A modular instrument}

To provide 32 simultaneously addressable integral fields of 4$''$
$\times$ 4$''$ covering a total 10-arcmin field, the instrument
consists of 8 unit spectrographs, as described above, each of which
has 4 movable arms and 4 image slicers.  The overall
dimensions of the instrument are 1.5\,m diameter and 2.5\,m long for
the packaging shown in Figure~5.  The mass of all the cold hardware is
700\,kg, and the total instrument mass will be about 2\,tonnes,
including cryostat, electronics, etc.

This very modular concept allows considerable freedom in the detailed
design. For example, it would be possible for one of the unit
spectrographs to have a different spectral resolution, or to dedicate
some arms to $K$-only spectroscopy and others exclusively to
$J$. Equally important, the modular design would enable a modular
development of the whole instrument.  To build and test one unit
spectrograph with 4 deployable fields is not a very big engineering
step from existing cryogenic IR spectrographs.  This unit could, of
course, be used for science while the others were being built.

\begin{figure}[b]
\begin{center}
\includegraphics[width=.7\textwidth]{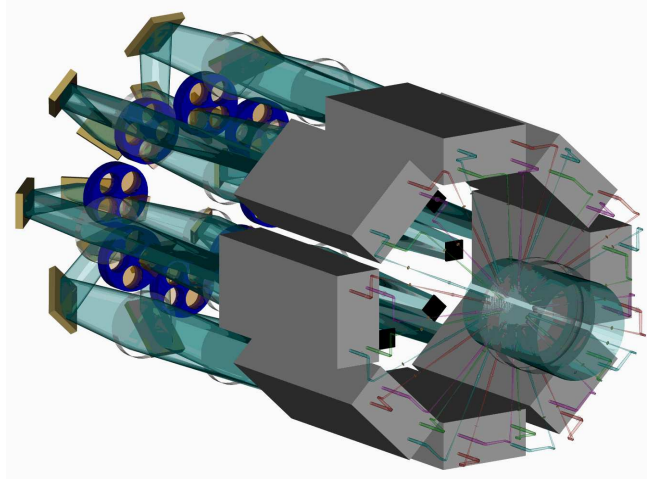}
\end{center}
\caption[]{8 unit spectrographs arranged around the focal plane provide 
32 deployable integral fields.}
\label{eps1}
\end{figure}

\section {Acknowledgements}
The instrument concept described here was based on design studies
carried out at the UK ATC and the University of Durham. Several people
participated or provided comments as the ideas progressed. We'd like
to thank Ian Parry, Eli Atad, Keith Taylor, Roger Haynes, Suzanne
Ramsay-Howatt, Ian Egan, Paul van der Werf, Antonio Chrysostomou, Arjun Dey 
for their scientific or technical contributions.

\end{document}